# PCA Methods and Evidence Based Filtering for Robust Aircraft Sensor Fault Diagnosis

N. Cartocci, G. Costante, M.R. Napolitano, P. Valigi *Member IEEE*, F. Crocetti, M.L. Fravolini

*Abstract—* In this paper PCA and D-PCA techniques are applied for the design of a Data Driven diagnostic Fault Isolation (FI) and Fault Estimation (FE) scheme for 18 primary sensors of a semi-autonomous aircraft. Specifically, Contributions-based, and Reconstruction-based Contributions approaches have been considered. To improve FI performance an inference mechanism derived from evidence-based decision making theory has been proposed. A detailed FI and FE study is presented for the True Airspeed sensor based on experimental data. Evidence Based Filtering (EBF) showed to be very effective particularly in reducing false alarms.

*Keywords—* Fault Isolation, Fault Estimation, PCA, D-PCA, Aircrafts, Evidence Based Filter.

## I. INTRODUCTION

Fault diagnosis deals with the problem of detecting and isolating sensors, actuators, and internal faults in dynamical systems. There are essentially two approaches to perform fault diagnosis: the model-based [1] and the data-based [2], [3] approaches. Model-based approaches use measured input-output signals and physical knowledge of the system to derive diagnostic signals, known as residuals, that are sensitive to faults [4]; while in data-based approaches diagnostic signals are derived directly from experimental data acquired from the system. Data-based approaches are preferable in case a detailed physical knowledge of the system is not available or when system input-output relations are too complex or uncertain [5]. Nowadays multivariate Statistical Process Monitoring (SPM) methods, particularly, Principal Component Analysis (PCA) [6] and Partial Least Squares (PLS) [7] are very popular tools for process monitoring. The main reason of this success lies in the ability of these techniques to manage efficiently large number of variables and large quantity of data; additionally, the mathematical formulation is straightforward, and the design effort is limited. These methods are widely used in industry and perform optimally in case of linearity and stationarity of the monitored system. Unfortunately, many industrial systems are characterized by dynamic responses, variable operating conditions and, more importantly, nonlinearities. These extremely relevant issues favored an intense research activity toward the development of advanced SPM methods [3].

N. Cartocci, G. Costante, P. Valigi, F. Crocetti, M.L. Fravolini are with the Department of Engineering, University of Perugia, Perugia, 06125 Italy (e-mails: nicholas.cartocci@studenti.unipg.it, gabriele.costantei@unipg.it, paolo.valigi@unipg.it, francesco.crocetti@unipg.it, mario.fravolini@unipg.it).

M.R. Napolitano is with the Department of Mech. and Aerospace Engineering, West Virginia University, Morgantown, WV, 26506, USA (e-mail: marcello.napolitano@mail.wvu.edu).

A relevant variation to the basic PCA and PLS methods is the Total Projection to Latent Structure (TPLS) approach proposed in [8] to handle the oblique decomposition problem of standard PLS and the concurrent projection to latent structures (CPLS) method [9]. The issue of system dynamics and temporal correlation in the data have been faced in variants such as the Dynamic PCA (D-PCA) [10], the dynamic PLS (DPLS) [11] and the Canonical Variate Analysis (CVA) approach [12]. More recently kernel-based PLS and PCA approaches have been proposed to address nonlinearity issues [13-14]. Other recent popular methods applied in industrial processes monitoring are based on the Fisher Discriminant Analysis (FDA) [15], and on the Independent Component Analysis (ICA) [16].

A well-known problem of many multivariate SPM approaches when used for Fault Isolation (FI) purposes is the so-called 'smearing effect', which essentially implies that the effect of a fault may propagate to other normal variables leading to a potential incorrect FI. To cope with this problem SPM methods are often used in conjunction with recursive Bayesian filtering that are used to propagate fault probability information through time with the purpose of improving FI reliability [17-19]. A possible problem of Bayesian approaches is that they require the knowledge of fault conditional probabilities to perform Bayesian inference. In practice, these conditional probabilities are extremely difficult to be computed and, for this reason, are often approximated with basic shapes whose parameters require a fine-tuning to perform satisfactory or are inferred from data.

In this study we propose an alternative inference approach derived from evidence based decision making theory [20]. This model is widespread for instance in Neuroscience where it is used to model the mechanisms of animal decision making in case of a Two-Alternative Forced Choice (TAFC) task [21-23].

In this paper basic PCA and D-PCA techniques, in conjunction with fault reconstruction [24] and Evidence Based Filtering (EBF), have been employed to design a robust data-based Fault Isolation scheme for 18 primary sensors of a P92 Tecnam semi-autonomous aircraft [25].

## II. PROCESS MONITORING BASED ON PCA MODELS

A PCA model is derived from a dataset of $m$ samples of a vector $x(k) \in \mathbb{R}^n$. The overall dataset is represented by a matrix $\mathbf{X} \in \mathbb{R}^{m \times n}$ defined as:

$$\mathbf{X} = [x(1), x(2), ..., x(m)]^T \qquad (1)$$

Data in (1) are scaled so that the $m$ signals in $\mathbf{X}$ have zero mean and unit variance. The covariance matrix is estimated from the normalized data and is computed as follows:

$$\mathbf{S} \cong (\mathbf{X}^T \mathbf{X})/(m-1) \qquad (2)$$

The PCA method performs the eigenvalue decomposition of the matrix $S$ in (2) to derive the principal and the residual loading matrices, $\hat{P} \in \mathbb{R}^{n \times l}$ and $\tilde{P} \in \mathbb{R}^{n \times (n-l)}$ respectively, where $l$ is the number of Principal Components (PCs) used in the model and $n-l$ is the number of Residuals components (Rs) [24]. The eigenvalue decomposition is such that

$$S = P\Lambda P^T = \begin{bmatrix} \hat{P} & \tilde{P} \end{bmatrix} \begin{bmatrix} \hat{\Lambda} & 0 \\ 0 & \tilde{\Lambda} \end{bmatrix} \begin{bmatrix} \hat{P} & \tilde{P} \end{bmatrix}^T \quad (3)$$

where the diagonal matrix $\hat{\Lambda}$ contains the principal eigenvalues and $\tilde{\Lambda}$ contains the residual eigenvalues; in addition, the overall transformation $P = [\hat{P} \ \tilde{P}]$ is orthonormal. Expanding (3), results

$$S = \hat{P}\hat{\Lambda}\hat{P}^T + \tilde{P}\tilde{\Lambda}\tilde{P}^T \quad (4)$$

A generic sample $x(k)$ can be decomposed as follows:

$$x = \hat{x} + \tilde{x} \quad (5)$$

where:

$$\hat{x} = \hat{P}\hat{P}^T x = \hat{C} x \quad (6)$$

is the projection of $x$ in the PCs Subspace and

$$\tilde{x} = \tilde{P}\tilde{P}^T x = \tilde{C} x \quad (7)$$

is the projection of $x$ in the Rs Subspace. Matrix $\hat{C} = \hat{P}\hat{P}^T$ is the projection matrix in the PCs subspace while $\tilde{C} = \tilde{P}\tilde{P}^T$ is the projection in the Rs subspace.

### A. The Dynamic PCA method

To take into account of possible autocorrelation in process variables, the Dynamic PCA (D-PCA) method has been proposed in [10]. In this approach the observations within a specified time interval [$k-d, k$] are considered within the same PCA model, where $d$ is the number on delayed samples considered in the "extended model". Let $x(k) \in \mathbb{R}^n$ be the collection of all variables at time $k$. In the D-PCA an extended variable vector is defined as

$$z(k) = [x^T(k), x^T(k-1)...x^T(k-d)]^T \in \mathbb{R}^{(n \cdot d) \times 1} \quad (8)$$

The associated extended matrix $Z \in \mathbb{R}^{m \times (n \cdot d)}$ is defined as

$$Z = [z(1), z(2), ..., z(m)]^T \quad (9)$$

The off-line design and the on-line computation are the same of standard PCA. That is the D-PCA is performed by simply substituting the data matrix $X$ in (1) with the extended data matrix $Z$ in (9) and repeating the decomposition steps (2-7).

### B. Fault detection indices

PCA-based fault detection techniques feature statistical indices for fault detection. The most popular indices are the Square Prediction Error (SPE) and the Hotelling's $T^2$ indices that are defined below.

*SPE Index*

SPE index is used to monitor the normal variability in the R$s$ subspace [24] and is defined as:

$$SPE = x^T \tilde{C} x = x_{\tilde{C}}^2 \quad (10)$$

The process is considered in normal operation if $SPE < \delta^2(\alpha)$ where the threshold $\delta(\alpha)$ is derived assuming a specified confidence level $(1-\alpha) \times 100$. A detailed discussion and control limit computation can be found in [26] under the Gaussian assumption of the data distribution.

*$T^2$ statistic*

The $T^2$ index measures the variation of a process in the PCs subspace [24] and it is defined as:

$$T^2 = x^T D x = x_D^2 \quad (11)$$

where $D = \hat{P}\hat{\Lambda}^{-1}\hat{P}^T$ is positive semidefinite. The process is in normal operation if $T^2 < \tau^2(\alpha)$, where $\tau$ is computed assuming a specified confidence level $(1-\alpha) \times 100$.

The SPE and $T^2$ indices have complementary properties and are often used jointly. A combined index based on SPE and $T^2$ indices has been proposed in [24].

## III. FAULT DIAGNOSIS BY CONTRIBUTION PLOTS

Following a fault detection, the subsequent step deals with the identification of the specific variable(s) that are responsible for the fault detection. A popular approach for fault isolation is the so-called Contributions Plot (CP) method [24]. The idea behind the CP method is that the variable providing the largest contribution to the fault detection index is the most probable faulty variable. Fault isolation is performed computing, at each sample time $k$, the contribution of all the variables to the detection index and assigning the fault to the variable having the largest contribution.

### A. SPE contributions

To better understand the contributions approach the SPE index is rewritten as:

$$SPE = \sum_{i=1}^{n} \left( \xi_i^T \tilde{C} x \right)^2 = \sum_{i=1}^{n} c_i^{SPE} \quad (12)$$

where

$$c_i^{SPE} = \left( \xi_i^T \tilde{C} x \right)^2 = \tilde{x}_i^2 \quad (13)$$

is the contribution of variable $x_i$ to the index SPE; the $\xi_i$ vector is the $i$-th column of the identity matrix and represents also the direction of the variable $x_i$.

### B. $T^2$ contributions

The contributions for the $T^2$ index are defined as

$$c_i^{T^2} = \left( \xi_i^T D^{0.5} x \right)^2 \quad (14)$$

Typical problems with basic contribution plots have been discussed by many authors, and can be summarized as:
1) Faults associated to a variable with a small contribution may not have the largest contribution unless the fault magnitude is large. This can be the cause of an incorrect FI
2) Contribution plots can lead to incorrect fault isolation regardless the fault amplitude even in the case of a simple (single) sensor fault. The above relevant limitations have motivated the development of alternative methods such as those discussed in the next section.

## IV. RECONSTRUCTION BASED CONTRIBUTIONS

Given an arbitrary fault direction, reconstruction based methods estimate the optimal fault amplitude of a potential additive fault along this direction such that a specific fault detection index is minimized [27]. In this study we have used the reconstruction-based approach proposed in [24] where possible fault contributions are reconstructed along the directions of the *n* monitored variables.

*A. Reconstruction based contributions of the RBC index*

The reconstructed vector along the *i-th* variable direction $\xi_i$ is defined as

$$z_i = x - \xi_i f_i \qquad (15)$$

The purpose of the reconstruction approach is to compute the minimum value for the scalar fault amplitude $f_i$ that minimizes the quadratic index $J = (x - \xi_i f_i)^T \tilde{C} (x - \xi_i f_i)$. It is simple to show that the $f_i$ minimizing $J$ is equal to:

$$f_i = (\xi_i^T \tilde{C} \xi_i)^{-1} \xi_i^T \tilde{C} x \qquad (16)$$

The reconstruction based contribution of the variable $x_i$ to the detection index *J* is defined as

$$RBC_i^{SPE} = \| \xi_i^T f_i \|^2 \qquad (17)$$

Substituting (16) into (17), after some computations, results in

$$RBC_i^{SPE} = \frac{(\xi_i^T \tilde{C} x)^2}{\tilde{c}_{ii}} = \frac{\tilde{x}_i^2}{\tilde{c}_{ii}} = \frac{c_i^{SPE}}{\tilde{c}_{ii}} \qquad (18)$$

Eq. (18) reveals that index $RBC_i^{SPE}$ is a scaled version of the index $c_i^{SPE}$. The important aspect is that the scaling coefficient is different for each fault direction; therefore, the $RBC_i^{SPE}$ and the $c_i^{SPE}$ indices are substantially different.
Although different, both indices suffer from the so-called 'smearing' effect, that is the fault in a variable produces a component also in the contributions plot associated to other variables. Clearly, this effect is undesirable since it may lead to incorrect fault isolations. Referring to the smearing effect, it was shown in [24] that under the assumption that the fault direction is exactly in the *i-th* sensor direction $\xi_i$, (that is $x = \xi_i f$), there is no guarantees that $c_i^{SPE} \geq c_j^{SPE}$ while is it guaranteed that

$$RBC_i^{SPE} \geq RBC_j^{SPE} \qquad (19)$$

Condition (19) is important because in case of a single sensor fault, the traditional contribution method does not guarantee correct fault isolation while the RBC is able to guarantees correct fault isolation. Although this result holds exactly only under the restrictive condition $x = \xi_i f_i$, it is highlighted that this is approximately true in case of large amplitude faults, that is when the fault free component of the signal is negligible with respect to the $\xi_i f_i$ component.

*B. Reconstruction based contributions of the T² index*

Using similar arguments of section IV.A it is possible to define the Reconstruction based contributions of the T² index as follows

$$RBC_i^{T^2} = \frac{(\xi_i^T D x)^2}{d_{ii}} \qquad (20)$$

Under the same hypothesis of condition (19) it can be shown that also the $RBC_i^{T^2}$ index guarantees correct fault diagnosis while this is not true in general for the contribution index $c_i^{T^2}$

*C. Fault modeling and fault isolation*

In this study, additive (single) failures $f_i(k)$ acting on a generic (*i-th*) sensor of $x(k)$ are considered, that is, the vector $x(k)$ is replaced, by its faulty version:

$$x(k) \leftarrow x(k) + \varepsilon_i f_i(k) \qquad (21)$$

where $\xi_i$ is the i-th column of the identity matrix **I**. We considered faults having a constant bias shape (step):

$$f_i(k) = \begin{cases} 0 & t < k_f \\ A_i & t \geq k_f \end{cases} \quad i = 1...n \qquad (22)$$

where $k_f$ is the fault time and $A_i$ is the fault amplitude. Following a fault detection, the Fault Isolation (FI) is performed (each sampling time *k*) associating the fault to the sensor producing the largest contribution index, that is

$$\text{Fault-index}(k) = \underset{i=1...n}{\mathrm{argmax}} \left( \text{Index}_i(k) \right) \qquad (23)$$

where $\text{Index}_i(k)$ is one of the contribution indices defined in (13), (14), (18) and (20).

## V. EVIDENCE-BASED FILTERING FOR ENHANCED FAULT ISOLATION

The FI methods described in Section IV builds on a decision strategy that isolates the faulty sensor using only the information of the current sample and does not consider the history of the fault index in previous instants. In this study, in order to account for temporal dependencies, a filtering approach based on evidence based decision theory [20] is proposed. This decision making approach is widespread, for instance, in neuroscience where it is used to explain the mechanisms of animal decision in case of Two-Alternative Forced Choice (TAFC) task [23]. TAFC models are typically based on three assumptions [23]: 1) evidence confirming each alternative is integrated over time; 2) the process is subject to random fluctuations; and 3) the decision is made when sufficient evidence has accumulated favoring one alternative over the other. We believe that these assumptions fit reasonably well with the FI problem.
Specifically, for FI purposes, we considered the so called Drift Diffusion Model (DDM) [23]. The DDM assumes that decisions are made by a noisy process that accumulates information over time from a starting point toward one of two response boundaries. Considering a DDM for a generic sensor, the state $s_i(k)$ represents the accumulated *difference* between the evidence supporting the two hypotheses (fault / no-fault on the *i-th* sensor). For each sensor, the state $s_i(k)$ evolves according to the DDM:

$$s_i(k+1) = s_i(k) + g_i \left( I^+(k), I^-(k), s_1(k)...s_m(k) \right) + n_i(k)$$
$$s_i(0) = s_{oi} \quad i = 1...n \qquad (24)$$

The functions $g_i(k)$ model the process of evidence accumulation and are known as the drift rates. These functions are determined by the information provided by the sensors readings that measure the evidence $I^+(k)$ supporting the hypothesis that the *i-th* sensor is faulty and on the evidence $I^-(k)$ supporting the hypothesis that $i$-th sensor is no-faulty. The $g_i(k)$ functions can, in general, depend also on the current level of accumulated evidence $s_j(k)$ j=1...n of the other sensors. The $n_j(k)$ term represents white noise.

In the present study a simple DDM was employed to characterize the evidence (belief) accumulation process to characterize the healthy/faulty status of the $i$-th sensor

$$s_i(k+1) = s_i(k) + g_i(k) \quad s_i(0) = 0 \quad (25)$$

where

$$g_i(k) = \begin{cases} +0.01 & \text{if sensor }(i)\text{ maximize the Index}(k) \\ -0.005 & \text{otherwise} \end{cases} \quad (26)$$

that is the isolated sensor (according to (21)) increases of 0.01 the evidence of $s_i(k)$ toward the positive direction (associated to the fault hypothesis) otherwise $s_i(k)$ is penalized by -0.005. The fault declaration threshold for $s_i(k)$ was set to 0.2 (that is, the optimistic case, the threshold is reached, after 20 steps). To keep the filter reactive, the $s_i(k)$ states in (25) is saturated above by 1 and below by 0. The rational of this approach is that only the more likely faulty sensor (according to the incoming measurements) receives a reward by increasing the accumulated evidence in the direction of the positive threshold. All the other sensors, instead, receive a negative reward that decreases the accumulated evidence of being faulty. Despite its simplicity, this approach showed to be robust and effective to limit wrong temporary fault isolations reducing the smearing effect especially in case of small amplitude faults.

## VI. AIRCRAFT AND FLIGHT DATA

The PCA-based techniques have been designed and validated using multi flight data of a Tecnam P92 aircraft, shown in Fig. 1. Data were acquired in semi-autonomous mode, that is the aircraft was manually flown by a pilot during the take-off, and landing phase and flown autonomously in cruise flight condition. A set of nine flight datasets was considered in this study of which five flights (2 hours and 20 minutes) were used for the design and the remaining four (2 hours and 21 minutes) for validation purposes; the data sampling time was 0.1 s. The study does not consider data associated with take-off, and landing phases. The considered 18 sensors are listed in Table-I.

## VII. PCA MODELS AND EXPERIMENTAL RESULTS ON VALIDATION FLIGHT DATA

The PCA-based techniques require the definition of the appropriate number $l$ of PCs to retain in model (3). Several strategies have been proposed in the literature to determine the correct number of components. In this study we have used a basic method that retains the minimum number of components that guarantee at least the 98% of explained variance in the PCs subspace of the training data. This leads to the selection of $l=14$ PCs for basic PCA methods and $l=23$ for D-PCA methods.

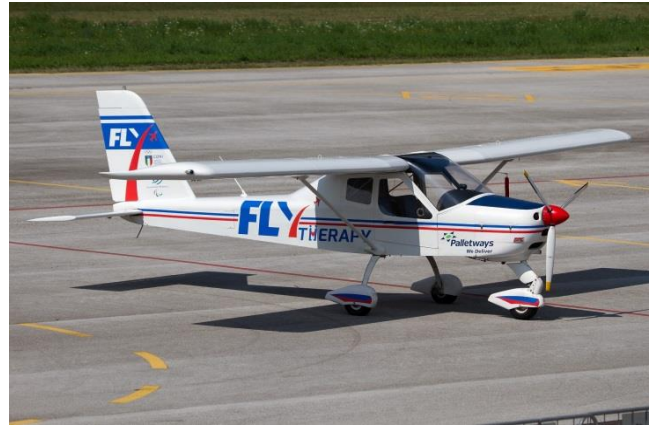

Fig.1 The semi-autonomous Tecnam P92 Aircraft.

TABLE I AIRCRAFT SENSORS

| α | Angle of attack | P | Roll speed | Zpos | GPS altitude |
|---|---|---|---|---|---|
| β | Drifting angle | Q | Pitch speed | Ap | Aileron position |
| TaS | True AirSpeed | R | Yaw speed | Rp | Rudder position |
| φ | Roll angle | NNx | Longitudinal load factor | Tp | Thrust lever position |
| θ | Pitch angle | NNy | Lateral load factor | Pp | PitchTrim position |
| ψ | Yaw angle | NNz | Vertical load factor | Sp | Stabilator position |

In this study, for space limitations, the analysis has been limited to the True Air Speed (*TaS*) sensor. However, it is underlined that the FI schemes consider the whole set of $n=18$ possibly faulty sensors, that is, the schemes may assign the fault to one among the 18 sensors. Further, since the main purpose of this research is to compare the performance of FI and FE techniques, the following analysis is performed assuming an "ideal" failure detection algorithm that detects the fault as soon as it is injected in a generic time instant $k=k_f$ (the fault detection delay is assumed equal to zero). The performance of the different techniques was evaluated and analyzed using the following two metrics.

Fault Isolation Percentage
Considering a fault on the $i$-th sensor of amplitude $A_i$ in the $j$-th validation flight, the Fault Isolation Percentage $I_{\%,i}$ is defined as the (percent ratio) between the number of samples the fault is correctly attributed to the i-th sensor and the total number of samples:

$$I_{\%\_i}(A_i) = 100 \cdot \sum_{j=1}^{N_{val}} N_{OK\_j,i}(A_i) / \sum_{j=1}^{N_{val}} N_{TOT\_j}(A_i) \quad (27)$$

where:
- $N_{val}$: Number of validation flights.
- $S_{I,j}$: Set of samples in validation flight-j from time $k_f$ until end of flight-j; $k_f$ is the fault injection time.
- $N_{OK\_j,i}$: For flight-j, this is the number of samples in $S_{I,j}$ that the algorithm correctly isolates the fault.
- $N_{TOT\_j}$: For validation flight-j, this is the total number of samples in the set $S_{I\_j}$.

Relative Fault Reconstruction Error.
The Relative Fault Reconstruction Error is defined as the percent ratio between the mean absolute fault amplitude reconstruction error for a fault on the $i$-th sensor, that is

$$E_{\%\_i}(A_i) = 100 \cdot \sum_{j=1}^{N_{val}} \sum_{k \in S_{I\_j}} \left| \frac{\hat{A}_{i\_j}(k) - A_i}{A_i} \right| / \sum_{j=1}^{N_{val}} N_{E\_j} \quad (28)$$

where:
- $A_i$ is the amplitude of a fault injected on the i-th sensor.
- $\hat{A}_i(k)$ is the estimated amplitude of the i-th fault.

The analysis was performed considering 100 equally spaced fault amplitudes $A_i$ in (22) in the range $[-A_{Mi}; A_{Mi}]$. The maximum amplitude was selected empirically so that for $A_i = A_{Mi}$ all the methods isolate correctly faults with index $I_{\%,i}$ at least equal to 70%. For the *TaS* sensor $A_M = 3$ m/s was fixed. The performance indexes were computed considering data from $N_{val}$=4 validation flights (2 h 21m). Table I reports, for sensor *TaS*, the Fault Isolation Percentage for the different techniques previously introduced (in the D-PCA methods it was considered a number of *d*=10 delays, therefore a time window of 1 s). In Table I the results are relative to small, medium, and large fault amplitudes. Table II reports the Relative Fault Reconstruction Error for the techniques based on the RBCs metric since these methods provides a direct estimation (through (16) $\hat{A}_i(k) = f_i$), of the fault amplitude.

TABLE I - FAULT ISOLATION PERCENTAGE [%] TAS SENSOR

| Fault amplitude | | Contributions + PCA | | RBCs + PCA | | Contributions + D-PCA | | RBCs + D-PCA | |
|---|---|---|---|---|---|---|---|---|---|
| | | SPE | $T^2$ | SPE | $T^2$ | SPE | $T^2$ | SPE | $T^2$ |
| TaS [m/s] | 1 | 29.65 | 0.45 | 21.77 | 41.84 | 5.87 | 3.52 | 42.52 | 87.12 |
| | 2 | 29.42 | 5.64 | 65.27 | 76.15 | 11.32 | 11.48 | 87.97 | 99.10 |
| | 3 | 34.41 | 13.61 | 89.91 | 90.03 | 29.30 | 12.64 | 97.12 | 99.91 |

TABLE II - RELATIVE FAULT RECONSTRUCTION ERROR [%] TAS SENSOR

| Fault amplitude | | RBCs + PCA | | RBCs + D-PCA | |
|---|---|---|---|---|---|
| | | SPE | $T^2$ | SPE | $T^2$ |
| TaS [m/s] | 1 | 12.64 | 32.11 | 13.18 | 12.63 |
| | 2 | 6.32 | 16.05 | 6.59 | 6.31 |
| | 3 | 4.21 | 10.70 | 4.40 | 4.21 |

Analyzing the results in the above tables the following conclusions can be drawn: the RBCs methods perform better than basic contribution plots methods (especially for large amplitude faults); the $T^2$ metric performs better than SPE metric; the inclusion of the delay in the D-PCA has, generally, a positive effect. Fig. 2 shows the FI performance (only for the RBCs methods) as function of the Fault Amplitude $A_i$. The analysis of the plot reveals that all the techniques are able to guarantee almost 100% correct FI for large enough fault amplitudes, confirming that the D-PCA techniques, in conjunction with $T^2$ index, provides the best FI performance. It is also observed that, for fault amplitudes smaller than 1 m/s, the % of correct FI is small. This occurs because in case of small fault amplitudes the modelling error has the same order of magnitude of the fault and, therefore, the FI is unreliable. Also in this case RBC and $T^2$ methods based on D-PCA provide the best performance. Fig. 3 shows the Relative Fault Reconstruction Errors for the RBC techniques. It is observed that for small fault, the fault amplitude estimate is inaccurate because the modeling error.

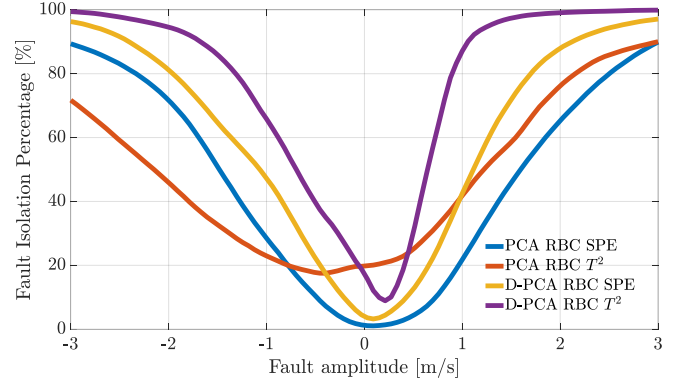

Fig. 2 - Fault Isolation Percentage for the TaS(k) sensor.

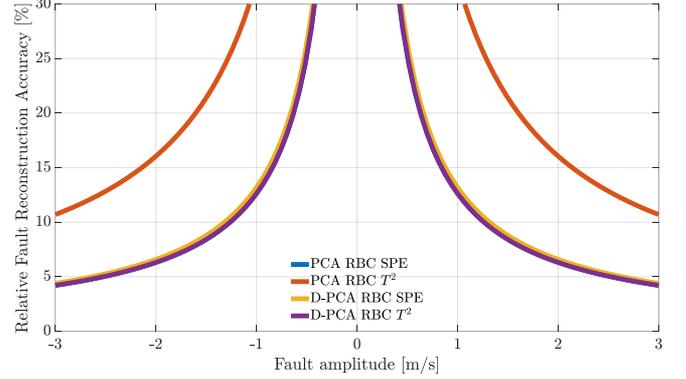

Fig. 3 - Relative Fault Reconstruction Accuracy for the TaS(k) sensor

Fig. 4 shows the Evolution of the sensor faulty Index computed using (23) for 20 minutes following the fault injection on the TaS sensor at $t_f$=0 (the EBF is not used) for a fault amplitude $A_f$=0.75 m/s in case of the D-PCA method with RBC $T^2$ metric. On the other side the results in case the EBF is active are shown in Fig. 5. It is evident that in Fig. 4 there are a significant number of periods where FI is incorrect, while in Fig. 5 the number of wrong FIs is significantly reduced thanks to the EBF. A quantitative analysis revels that the Fault Isolation Percentage without EBF filtering is approximately 67%, while with EBF this about 94%. Finally, Fig. 6 shows the evolution of the EBF outputs $s_i(k)$, $i$=1…18 for the first 60 seconds following the fault. It is evident that, following a transient, the accumulated evidence for the TaS sensor becomes dominant in approximately 5 s and maintains over the threshold (0.2) for the remaining portion of the flight.

## VIII. CONCLUSIONS

A data-driven Fault Isolation scheme for 18 primary sensors of a semi-autonomous Aircraft based on PCA techniques has been presented. The study confirms that Reconstruction Based Contributions method provides superior performance compared to the simpler Contributions Based method. The performance of Dynamic-PCA methods are remarkably superior to basic PCA conforming that the temporal correlation is important for this study. Experiments show that for the D-PCA method it is possible to achieve 99% correct FI for fault amplitude larger than 2 m/s on the True air Speed sensor evaluated on 4 validation flights. The employment of Evidence based Filtering applied to the RBC indexes showed to be very effective in further reducing wrong fault isolations during the flight.

ACKNOWLEDGMENT

This work was funded by University of Perugia 2017 basic research grant RICBA17MFR and by Clean Sky 2 Joint Undertaking (CS2JU), EU Horizon 2020 research and innovation program, under project no. 821079, E-Brake.

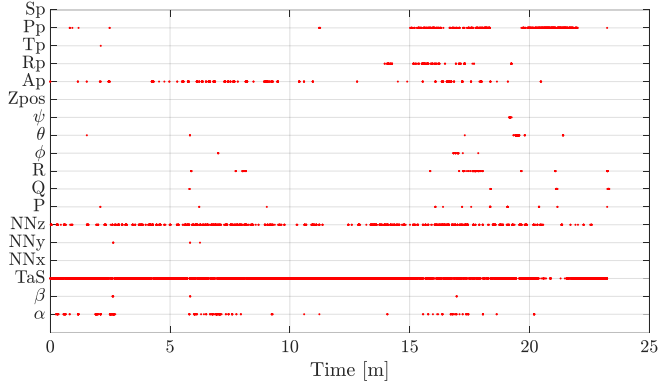

Fig. 4 – Evolution of the faulty-declared sensor without EBF

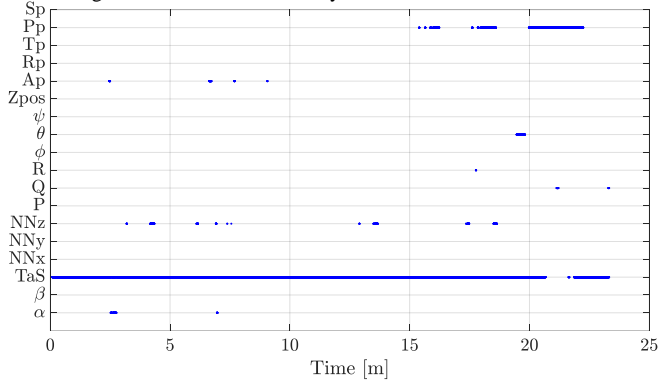

Fig. 5 – Evolution of the faulty-declared sensor with EBF

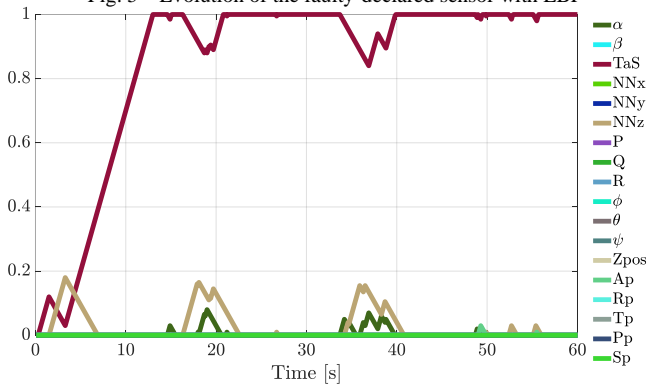

Fig. 6 – Evolution of the belief $s_i(k)$ in the first 60 seconds

REFERENCES

[1] S. X. Ding, *Model-based fault diagnosis techniques: Design schemes, algorithms, and tools*. Springer-Verlag, 2008.
[2] S. X. Ding, *Data-driven Design of Fault Diagnosis and Fault-tolerant Control Systems*. London: Springer London, 2014.
[3] S. Yin, S. X. Ding, X. Xie, and H. Luo, "A review on basic data-driven approaches for industrial process monitoring," *IEEE Trans. on Industrial Electronics*, vol. 61, no. 11. pp. 6414–6428, 2014.
[4] R. Isermann, *Fault-diagnosis systems : an introduction from fault detection to fault tolerance*. Springer, 2006.
[5] M. Da Ma, D. S. H. Wong, S. S. Jang, and S. T. Tseng, "Fault detection based on statistical multivariate analysis and microarray visualization," *IEEE Trans. Ind. Inf.*, vol. 6, no. 1, pp. 18–24, 2010.
[6] J. Gertler and J. Cao, "PCA-based fault diagnosis in the presence of control and dynamics," *AIChE J.*, vol. 50, no. 2, pp. 388–402, 2004.
[7] G. Li, S. J. Qin, and D. Zhou, "Geometric properties of partial least squares for process monitoring," *Automatica*, vol. 46, no. 1, pp. 204–210, 2010.
[8] D. Zhou, G. Li, and S. J. Qin, "Total projection to latent structures for process monitoring," *AIChE J.*, vol. 56, no. 1, pp. 168-178, 2009.
[9] S. J. Qin and Y. Zheng, "Quality-relevant and process-relevant fault monitoring with concurrent projection to latent structures," *AIChE J.*, vol. 59, no. 2, pp. 496–504, 2013.
[10] W. Ku, R. H. Storer, and C. Georgakis, "Disturbance detection and isolation by dynamic principal component analysis," *Chemom. Intell. Lab. Syst.*, vol. 30, no. 1, pp. 179–196, 1995.
[11] S. W. Choi and I. B. Lee, "Multiblock PLS-based localized process diagnosis," *J. Proc. Control*, vol. 15, no. 3, pp. 295–306, 2005.
[12] Y. Wang, D. E. Seborg, and W. E. Larimore, "Process monitoring based on canonical variate analysis," in *ECC 1997 - European Control Conference*, 1997, pp. 3089–3094.
[13] R. Rosipal, L. J. Trejo, N. Cristianini, J. Shawe-Taylor, and B. Williamson, "Kernel Partial Least Squares Regression in Reproducing Kernel Hilbert Space," *J, Machine Learning Res*. vol. 2, no.12, pp. 97-123, 2001.
[14] C. F. Alcala and S. J. Qin, "Reconstruction-Based Contribution for Process Monitoring with Kernel Principal Component Analysis," *Ind. Eng. Chem. Res.*, vol. 49, no. 17, pp. 7849–7857, 2010.
[15] G. J. McLachlan, *Discriminant analysis and statistical pattern recognition*. Wiley, 1992.
[16] J. M. Lee, C. K. Yoo, and I. B. Lee, "Statistical process monitoring with independent component analysis," *J. Process Control*, vol. 14, no. 5, pp. 467–485, 2004.
[17] Y. Zheng, S. Mao, S. Liu, D. S. H. Wong, and Y. W. Wang, "Normalized Relative RBC-Based Minimum Risk Bayesian Decision Approach for Fault Diagnosis of Industrial Process," *IEEE Trans. Ind. Electron.*, vol. 63, no. 12, pp. 7723–7732, 2016.
[18] A. Haghani, T. Jeinsch, M. Roepke, S. X. Ding, and N. Weinhold, "Data-driven monitoring and validation of experiments on automotive engine test beds," *Control Eng. Pract.*, vol. 54, pp. 27–33, 2016.
[19] J. Liu, D. S. H. Wong, and D. S. Chen, "Bayesian filtering of the smearing effect: Fault isolation in chemical process monitoring," *J. Process Control*, vol. 24, no. 3, pp. 1–21, 2014.
[20] V. V. Baba and F. HakemZadeh, "Toward a theory of evidence based decision making," *Manag. Decis.*, vol. 50, no. 5, pp. 832–867, 2012.
[21] R. Bogacz, E. Brown, J. Moehlis, P. Holmes, and J. D. Cohen, "The physics of optimal decision making: A formal analysis of models of performance in two-alternative forced-choice tasks," *Psychol. Rev.*, vol. 113, no. 4, pp. 700–765, 2006.
[22] S. Tajima, J. Drugowitsch, and A. Pouget, "Optimal policy for value-based decision-making," *Nat. Commun.*, vol. 7, Aug. 2016.
[23] R. Ratcliff and G. McKoon, "The diffusion decision model: Theory and data for two-choice decision tasks," *Neural Computation*, vol. 20, no. 4. pp. 873–922, 2008.
[24] C.F. Alcala and S.J. Qin, "Reconstruction-based contribution for process monitoring," *Automatica*, vol. 45, no. 7, pp. 1593–160, 2009.
[25] M. L. Fravolini, G. Del Core, U. Papa, P. Valigi, and M. R. Napolitano, "Data driven schemes for robust Fault Detection of Air Data System Sensors," *IEEE Transation Control Syst. Technol.*, vol. 27, no. 1, pp. 234–248, 2017.
[26] S. J. Qin, "Survey on data-driven industrial process monitoring and diagnosis," *Annu. Rev. Control*, vol. 36, no. 2, pp. 220–234, 2012.
[27] R. Dunia and S. Joe Qin, "Subspace approach to multidimensional fault identification and reconstruction," *AIChE J.*, vol. 44, no. 8, pp. 1813–1831, 1998.